\documentstyle[preprint,aps]{revtex}
\begin{document}
\draft
\preprint{NT@UW-98-20}
\title{ 
Rotational Invariance In Nuclear  Light-Front Mean Field Theory}

\author{P. G. Blunden}
\address{Department of Physics and Astronomy\\University of Manitoba\\
Winnipeg, MB, Canada R3T 2N2}

\author{M. Burkardt}
\address{Department of Physics\\
New Mexico State University\\
Las Cruces, NM 88003-0001, U.S.A.}

\author{G. A. Miller}
\address{Department of Physics, Box 351560\\
University of Washington \\
Seattle, WA 98195-1560, U.S.A.}
\date{\today}
\maketitle
\begin{abstract}
  
A light-front treatment for spherical nuclei
is developed from a relativistic
effective
Lagrangian and employing the mean field approximation. Minimizing the nuclear
minus momentum subject to the constraint that, in the rest frame,
the expectation values of the
plus- and minus-momenta are the same leads to a formalism in which rotational
invariance is recovered.

\end{abstract}
\narrowtext

Light cone variables have long been used to analyze high energy experiments
involving nuclear and nucleon targets. The most well-known use appears in the
 parton model. The Bjorken variable $x$ is the
 ratio $x=k^+/p^+$ where  $k^+=k^0 +k^3$ is the plus-momentum of the struck
 quark and $p^+$ is the plus-momentum of the target. Quark distributions
 represent the probability that a quark has a plus-momentum fraction $x$.
 Our focus here is
 on the distribution functions which describe the plus-momentum
 carried by 
 the nucleons within the nucleus. Such distributions, which
 depend on $p^+$ of the struck nucleon, are needed to
 analyze deep inelastic lepton-nucleus scattering\cite{emcrevs}, and
 also enter
 in the analysis of high-momentum transfer
 nuclear quasi-elastic reactions such as  $(e,e'), (e,e'p),(p,2p)$
 \cite{west}-\cite{jf}. 
One may also obtain similar distributions for the
 nuclear mesons, and these enter in the analyses of nuclear
 Drell-Yan experiments\cite{dyth}. If the light front formalism is used, these
 distributions are simply related to the square of the nuclear   ground
 state wave functions\cite{lcrevs}. Many nuclear  high momentum transfer
 experiments
 are planned, so that it is necessary to derive a relativistic formulation
 which uses $p^+$ variables closely related to experiments, and which
 incorporates the full knowledge of nuclear dynamics. 

 If  
 the relevant  nuclear wave functions  depend on $p^+$, the 
 canonical spatial variable  is $x^-=x^0-x^3$. This leaves  
 $x^+=x^0+x^3$ to be  used as a
 time variable, with  the light front Hamiltonian ($x^+$ development operator)
 as $P^-=P^0-P^3$. These are the light front variables of Dirac\cite{Dirac}.
  There are clear  advantages 
in using these
variables,  but
a principal problem arises because the use of $x^+$ as ``time'' and
$x^-,\bbox{x}_\perp$ as ``space'' involves the loss of manifest rotational
invariance. This is especially important in nuclear physics because the
understanding  of magic numbers rests on the $2j+1$ degeneracy of single
particle orbitals of good angular momentum.
Our  purpose  is to show how rotational invariance 
is recovered in nuclear  mean field theory. Once this is accomplished,
 light front
dynamics can   be a useful tool in a variety of
nuclear physics calculations.

We  start by assuming a field-theoretic Lagrangian
in which the  
nuclear constituents are nucleons $\psi$ 
 scalar mesons $\phi$ and
 vector mesons
$V^\mu, V^{\mu\nu}=\partial^\mu V^\nu-\partial^\nu V^\mu$. 
The  Lagrangian 
is a standard one\cite{bsjdw},  and
we use 
the meson-nucleon coupling constants and meson masses
of Horowitz and Serot\cite{hs}.  The use of  
 light front dynamics  allows
us to compute the necessary nuclear
meson distribution functions using the same formalism that
yields the nuclear binding energy and density.
Previously,   one of us  constructed a light 
front mean-field treatment of nuclear matter\cite{jerry} using the same
Lagrangian.  
Vector mesons were found to  carry 35\% of the
total plus-momentum (mass), with  each vector meson carrying
an infinitesimally
small amount of plus-momentum. Does 
 such a result hold for finite nuclei?


 Light front quantization of a given Lagrangian
is a standard procedure\cite{lcrevs}.
 One first uses the Lagrangian to
 derive field equations which enable the   identification of
  dependent degrees and their elimination in favor of
 independent ones. The procedure for the present Lagrangian
 is described elsewhere \cite{jerry},
 so we discuss only the most relevant features
concerning  the nucleons. 
Two of the four degrees of freedom
of this  spin 1/2 particle
 entering in the usual Dirac spinor
must be redundant.  These
 are separated using  the 
projection operators
$\Lambda_\pm\equiv {1\over 2}\gamma^0\gamma^\pm$,  with
$\psi_\pm\equiv\Lambda_\pm \psi$.
The independent fermion degree of freedom
is chosen as   $\psi_+$. In general  the
equation for $\psi_-$ depends on the interaction $V^+$ in a complicated
manner. This dependence is eliminated by using the transformation
of Refs.~\cite{des71}. The net result is that a 
   transformed vector meson field
$ \bar V^\mu=V^\mu-{\partial^\mu\over \partial^+} V^+
 \label{vbar}$
enters in the nucleonic field equation, while $V^{\mu}$ enters in the
equation for the vector meson field.

The energy momentum tensor is $T^{\mu\nu}$, and the 
volume integrals of components $T^{+-}$ and  $T^{++}$ are the
Hamiltonian
$P^-$ and  momentum  $P^+$. We find
\begin{eqnarray}
T^{+-}&=&
(\bbox{\nabla}_\perp\phi)\cdot(\bbox{\nabla}_\perp\phi) +m_s^2\phi^2 -
(\bbox{\nabla}_\perp V^+)\cdot(\bbox{\nabla}_\perp V^-) -m_v^2 V^+ V^-
+2\psi^\dagger_+\left(i {\partial}^-
-g_v\bar V^-\right)\psi_+,\label{tpm1}\nonumber\\
T^{++}&=&\partial^+\phi \partial^+\phi +
(\bbox{\nabla}_\perp V^+)\cdot(\bbox{\nabla}_\perp V^+) +m_v^2 V^+ V^+
+2\psi^\dagger_+ i\partial^+ \psi_+\;.
\label{tpp}
\end{eqnarray}
We choose to work in the nuclear rest frame so that
$ P_\perp\mid\Psi\rangle = 0$ and
$ P^+\mid\Psi\rangle = M_A\mid\Psi\rangle.$
In this frame, the light-front
Schr\"odinger equation for the ground-state
wave function $\mid\Psi\rangle$ reads
\begin{equation}
 P^-\mid\Psi\rangle = \frac{M_A^2+P_\perp^2}{P^+} 
\mid\Psi\rangle= M_A\mid\Psi\rangle.
\end{equation}
In light-front dynamics the goal is to solve the equation
$\left(P^+ P^--P_\perp^2\right)\mid\Psi\rangle = 
M_A^2\mid\Psi\rangle $ with fixed values for $P_\perp$ and $P^+, $
using a Fock-space basis. This is not possible for  systems containing
many nucleons, so we need to construct a light front version of the
Hartree-Fock procedure in which the energy of a Slater determinant is
minimized. 
Minimizing  the expectation value of $P^-$ doesn't work because
one  gets a zero  of $P^-$ for an infinite value of 
$P^+$. Instead we  minimize the expectation value of
$P^-$ subject to the condition that the expectation
value of $P^+$ is equal to
the expectation value of $P^-$.
This is the same as
minimizing the average of $P^-$ and $P^+$.
To this end
we define a light front Hamiltonian:
$
  H_{LF}\equiv {1\over 2}\left(P^++P^-\right)$.
This is not  the usual Hamiltonian because 
light-front quantization
is used to define all of the operators that enter.

The wave function $\mid\Psi\rangle$ consists of a
Slater determinant of nucleon
fields $\mid\Phi\rangle$ times a mesonic portion,
and the mean field approximation is characterized by the replacements:
$
  \phi\to \langle\Psi\mid \phi\mid\Psi\rangle,
V^\mu\to \langle\Psi\mid V^\mu\mid\Psi\rangle. 
$
We  quantize the
nucleon   fields using
\begin{equation}
\psi(x) =\sum_n \langle x^-,x_\perp\mid n\rangle e^{-i p_n^-x^+/2}\;b_n\;,
\end{equation}
in which the variable $x$ represents both the spatial variables
$\bbox{x}_\perp,x^-$ and the spin, isospin indices, and $b_n$ obeys the usual
anti-commutation relations. The meson fields are also functions of
$ x^-,x_\perp$. The  single particle
states $\mid n\rangle$ are defined by minimizing the nucleonic contribution to
the expectation value
of $\langle\Psi\mid H_{LF}\mid\Psi\rangle$.
 The Slater
determinant $\mid\Phi\rangle$ is defined by allowing $A$ nucleon states to be
occupied.
The results of the minimization are
 \begin{eqnarray} 
 p_n^- \mid n\rangle_+&=&\left(i\partial^++
2 g_v\bar{V}^-\right)
 \mid n \rangle_+ +
 (\bbox{\alpha}_\perp\cdot 
(\bbox{p}_\perp-g_v\bbox{\bar V}_\perp)+\beta(M+g_s\phi))\mid n \rangle_-\;,
\nonumber\\
 i\partial^+\mid n\rangle_-&=& \left(\bbox{\alpha}_\perp\cdot 
(\bbox{p}_\perp-g_v\bbox{\bar V}_\perp)+\beta(M+g_s\phi)\right)\mid n
\rangle_+\;,
 \label{nplus}
\end{eqnarray}
where $\mid n\rangle_\pm=\Lambda_\pm\mid n\rangle$.
Minimization of  $\langle\Psi\mid H_{LF}\mid\Psi\rangle$
with respect to the meson
field variables leads to the  meson field equations. We  find 
     \begin{eqnarray}
       \left(-\nabla_\perp^2-
         \left(2{\partial \over \partial x^-}\right)^2+m_s^2 
       \right)
        \langle \Psi\mid \phi(x)\mid \Psi\rangle &=& -g_s \rho_s(x),
        \label{ph1}\\
  \left(-\nabla_\perp^2-
         \left(2{\partial \over \partial x^-}\right)^2+m_s^2 
       \right)     
        \langle \Psi\mid {V}^\mu (\bbox{r})
        \mid \Psi\rangle &=& g_v j_B^\mu(x),
        \label{veq}
\end{eqnarray}
where $ \rho_s(x)\equiv\langle \Psi\mid 
        {\bar\psi}(x)\psi(x)\mid \Psi\rangle$ and
        $ j_B^\mu(x)\equiv\langle \Psi\mid
        \bar\psi(x)\gamma^\mu\psi(x)\mid \Psi\rangle$.

We next discuss how Eq.~(\ref{nplus}) is  solved.
The first step is to use a representation\cite{harizhang}
in which $\mid n\rangle_\pm$ are each represented as Pauli spinors.
The light front quantization respects 
manifest rotational invariance for rotations 
about the $z$-axis. Thus each single-particle state has a good $J_z$. 
We use a momentum representation for the longitudinal variable
in which the values of $p^+$ take on those
of a discrete set:
$p_m^+=(2m+1)\pi/(2L)$ where $m\ge0$ and $L$ is a quantization
length. 
This means that the nucleon  wave functions
have support only for  $p^+\ge0$. This spectrum condition is
a requirement for exact solutions for  any theory.
The coordinate space representation
is used for the $\perp$ variables.
Then we 
have 
\begin{eqnarray}
\langle p_m^+,\bbox{x}_\perp\mid n\rangle_+=\left[\begin{array}{c}
U_m^{(n)}(x_\perp)e^{i(J_z-1/2)\phi}\\
L_m^{(n)}(x_\perp)e^{i(J_z+1/2)\phi}\end{array}
\right],\label{spinor}
\end{eqnarray}
in which  the upper (lower) entrees of the Pauli spinor correspond to 
$m_s=1/2(-1/2)$, and $\bbox{x}_\perp\equiv (x_\perp,\phi)$.
When Eq.~(\ref{spinor}) is used in Eq.~(\ref{nplus}),
one finds that the equations 
do not depend on the magnitude of $J_z$;  solutions for $\pm$ 
a given magnitude of $J_z$ are degenerate.
The functions $U_m(x_\perp)$ and $L_m(x_\perp)$ are expanded in a basis
of B-splines of degree five in $x_\perp$ \cite{bspline}.
One starts with a  guess for the
scalar and vector potentials. Then one solves Eq.~(\ref{nplus}). 
The solutions are used  to generate new scalar and vector nucleon
densities, which in turn  generate new potentials
via Eqs.(\ref{ph1}) and (\ref{veq}). This iteration  procedure is 
completed until the solutions are stable.

If these  solutions are to have any relevance at all,  they should
respect rotational invariance. The success in achieving this
is examined in Table I, which gives our results for the spectrum of $^{40}$Ca.
We see that the 
$J_z=\pm1/2$ spectrum contains the eigenvalues of all states, since all
states must have a $J_z=\pm1/2 $ component. Furthermore, the
essential feature that
the expected degeneracies among states with different values of $J_z$
are reproduced numerically.

The results shown in  Table 1 are obtained using 
a basis of 20 splines, a  box size of $2L = 24$ fm,
and 24 Fourier components in the expansion of the wavefunction.
 This value of  $L$ is  large enough so that our results do not depend
on it, and the 
 number of terms in the  expression for the density is enough to
ensure that the densities are spherically symmetric\cite{sphere}.
Another feature is that the spectrum with $p^+ > 0$ has no
negative energy states, 
so that in using the LF method one is  working in a basis of positive
energy states only.

The single particle energies $\epsilon_n$ reported in Table 1
 are given  by $\epsilon_n = p_n^-/ 2$.
The values  of $\epsilon_n$ are essentially
the same as those of the ET formalism to within the expected
numerical accuracy of our program, as are the nuclear densities. This
equality is not mandated by spherical symmetry alone because the solutions
in the equal time framework have non-vanishing 
components  with  negative values of  $p^+$. We may understand the 
near equality of single particle eigenvalues using an analytic argument.
First we use a representation in which $V^\mu$ appears in the equation
for the nucleon field. That is, define
$\langle x\mid n\rangle'\equiv e^{-ig_v\Lambda(x)}\langle x\mid n\rangle$,
with $\partial^+\Lambda=V^0$.
Then multiply the equation for $\mid n\rangle'_+$ by 
 $\gamma^+$ and the  equation
 for $\mid n\rangle'_-$  by $\gamma^-$. Adding the resulting two equations
  gives
\begin{equation}
(\gamma^0(p^-_n-2\gamma^0g_vV^0
-\gamma^3(2p^+-p^-_n/2))\mid {n}\rangle'
=2(\bbox{\gamma}_\perp\cdot  \bbox{p}_\perp+M+g_s\phi  
)\mid {n}\rangle'.\label{almost}
\end{equation}
Solving Eq.~(\ref{almost}) in coordinate space  is expected to lead to
solutions in which the 
spectrum condition is not respected exactly. Equation~(\ref{almost})
may be converted into a manifestly rotationally invariant
equation using the relation:
$x^-=- 2z,$ so
that $p^+=i\partial^+=
2i{\partial\over \partial x^-}\to-i{\partial\over \partial z}$.
The operator $p^+$ acts as a $p^3$ operator, and the
result (\ref{almost}) looks like the Dirac equation of the equal time (ET)
formulation (of eigenvalue $p^-_n/2$)
except for an offending term  $-p^-_n/2$
multiplying the $\gamma^3$. This term may be eliminated by including a phase 
factor: 
$\langle x\mid n\rangle'=e^{ip^-_n z/2}\langle x\mid n\rangle_{ET}.$
The result is that $\langle x\mid n\rangle_{ET}$ satisfies the standard
Dirac equation of the equal time formulation. This means that
the eigenvalues $\epsilon_n$ must be approximately the same in the two formulations.
The net result of these transformations is that if one were to include
$p^+<0$ in the LF calculation then the eigenvalues $\varepsilon_n$
in the LF and ET calculations would be identical and the LF
wave function would be expressed as
\begin{equation}
\langle p^+,x_\perp\mid n\rangle={1\over\sqrt{2\pi}}\int_{-\infty}^\infty dz
e^{-i(p^+-p_n^-/2)z}e^{ig_v\Lambda(x_\perp,z)}
\langle z,x_\perp\mid n\rangle_{ET}. \label{emcv}
\end{equation}
The 
relation (\ref{emcv}) tells us that 
$\langle p^+,x_\perp\mid n\rangle$ peaks at $p^+\approx M-g_VV^0$, with a width
of the order of the inverse of the radius of the entire nucleus. 
Therefore, neglecting $p^+<0$ is only a minor approximation, which
is the reason why our above LF calculation, with $p^+<0$ 
suppressed, leads to eigenvalues $\varepsilon_n$ that are 
approximately equal to the ET results.

One can also read off from Eq. (\ref{emcv}) that
the influence of the vector 
potential is to remove plus-momentum from the nucleons. Furthermore, the large
value of the nuclear radius causes the region of support to be very narrow, so
that $\langle p^+,x_\perp\mid n\rangle$ is very small for negative values of
 $p^+$.
 
The nuclear
energy is obtained by adding the
contributions to the expectation value of $P^-+P^+$
from the nucleons, and scalar and vector mesons.  Then
straightforward calculation leads to the result
 \begin{equation}
   M_A=\sum_{n<F}\epsilon_n 
   -{1\over 2}\int d^3r\left[g_s\phi(r)\rho_s(r)+ g_vV^0(r)\rho_B(r)\right],
\end{equation}
 where $\bbox{r}\equiv(\bbox{x}_\perp,z=-x^-/2)$.
The above expression is  the  result obtained in
standard equal time calculations, and the  nuclear binding energy we
obtain is  essentially
the same as that of Ref.~\cite{hs}.

The previous results mean
that we have reproduced the standard good results for nuclear
binding energy and density using a formalism that allows us to compute
nucleon and meson 
plus-momentum distributions  used to analyze various high energy
nuclear experiments.
   The probability that a nucleon $f_N(p^+)$   has a momentum $p^+$  
   is given simply by
    \begin{equation}
   f_N(p^+)
   =\sum_{n<F}
   \int d^2x_\perp \mid\langle{\bf x}_\perp,p^+\mid n\rangle\mid^2.
   \end{equation}
It is convenient to express the distribution in terms of the
dimensionless variable $y\equiv p^+/(M_A/A)$.
The result is shown in Fig.~1. The peak of the distribution is at
$y\approx 0.76$, corresponding to $p^+$ of 710~MeV
(which is the plus-momentum per nucleon).
Our numerical calculations show that the average value of $p^+$ is
672.56 MeV ($<y>=0.72$) out of a total expectation value $P^+/A$ of 930.56 MeV.
($P^+/A$ is the total energy per nucleon.)
The distribution is not symmetric about its average value, as it would
be if a simple Fermi gas model were used. Both of
these effects are caused by the
presence of nuclear mesons, which carry the remainder of the plus-momentum.

The meson distributions are computed using 
  \begin{equation}
f_s(k^+)=\int d^2k_\perp
\mid\langle\Psi\mid a({\bf k})\mid\Psi\rangle\mid^2 \label{fs}
     \end{equation}
for the scalar meson distribution, and 
   \begin{equation}
f_v(k^+)=\int d^2k_\perp \sum_{\omega=1,3}
\mid\langle\Psi\mid 
a({\bf k},\omega)
\mid\Psi\rangle\mid^2\label{fv}
\end{equation}
for 
the vector  meson distribution. The operator
$a({\bf k})$ destroys a scalar meson of momentum $k=(k^+,k_\perp)$,
and $a({\bf k},\omega)$ destroys a vector meson of polarization $\omega$.
The mean field approximation is used here, and the matrix elements needed
to evaluate Eqs.~(\ref{fs}) and (\ref{fv}) can be obtained using  standard
manipulations on Eqs.~(\ref{ph1}) and (\ref{veq}).
The detailed expressions for $f_s,f_v$
have   the same form as those of  Ref.~\cite{bm98}
 in which the nucleus was treated as an elementary
 spherical source of scalar and vector
 mesons.   The difference is that self-consistent field theory is used here
 to obtain the nucleon densities $\rho_{s,B}$.
The scalar mesons are found to carry  4.70 MeV of the plus-momentum,
while the vector mesons carry  253.30 MeV (or 27\%) of the nuclear
plus-momentum.  The technical reason for the
difference is that the evaluation of
$a^\dagger({\bf k},\omega)a({\bf k},\omega)$
counts vector mesons ``in the air''and the resulting expression contains 
 polarization vectors that give factor of ${1\over k^+}$ which enhance the
 distribution of vector mesons of low $k^+$.
 The results for the vector meson distribution are shown in Fig.~2.

It is worthwhile to see how  the present results are related to  
 lepton-nucleus deep inelastic scattering experiments. We find that the
 nucleons carry only 72\% of the plus-momentum. The use of our
 $f_N$ in standard convolution formulae lead to a reduction in the nuclear
 structure function that is far too large ($\sim$95\% is needed\cite{emcrevs})
 to account for the  reduction
 observed\cite{emcrevs} in the vicinity of $x\sim 0.5$. The reason for this
 is that the quantity $M +g_s\phi$ acts as a nucleon effective mass of about
 670 MeV, which is very small. A similar difficulty occurs in the $(e,e')$
 reaction  
\cite{frank} when the mean field theory is used for the initial and final
states. The use of a small effective mass and a large vector potential
enables a simple reproduction of the nuclear spin orbit force\cite{bsjdw}.
However, effects beyond the mean field may lead to a significant effective
tensor coupling of the isoscalar vector meson\cite{ls} and to an increased
value of the effective mass. Such effects are incorporated in Bruckner theory
and a light front version \cite{rmgm98} could be applied to finite nuclei with
better success in reproducing the data.

In any case, these kinds of nuclear physics calculations can be
done in a manner in which modern nuclear dynamics is respected,
boost invariance in the $z$-direction is preserved,  
and in which the rotational invariance so necessary to understanding the basic
features of nuclei is maintained.

\begin{table}[t]
\caption{Comparison of the single particle spectra of $^{40}$Ca in the
equal time (ET) formalism ($\epsilon_n-M_N$) with the light front (LF)
method ($p_n^-/2-M_N$).  Scalar and vector meson parameters are taken from
Horowitz and Serot~[9], and we have assumed isospin symmetry.}
\rule{0in}{2ex}
\begin{center}
\begin{tabular}{ldddd}
\multicolumn{2}{c}{ET} & \multicolumn{3}{c}{LF} \\
\cline{1-2} \cline{3-5}
State $n$ & $\epsilon_n-M_N$ (MeV) & $J_z=1/2$ & $J_z=3/2$ & $J_z=5/2$\\
\cline{1-1} \cline{2-2} \cline{3-3} \cline{4-4} \cline{5-5}
0s$_{1/2}$ & $-$55.40 & $-$55.39 & & \\
0p$_{3/2}$ & $-$38.90 & $-$38.90 & $-$38.90 & \\
0p$_{1/2}$ & $-$33.18 & $-$33.18 & & \\
0d$_{5/2}$ & $-$22.75 & $-$22.75 & $-$22.75 & $-$22.74 \\
1s$_{1/2}$ & $-$14.39 & $-$14.36 & & \\
0d$_{3/2}$ & $-$13.87 & $-$13.87 & $-$13.88 & \\
\end{tabular}
\end{center}
\label{table1}
\end{table}

\newpage
\begin{figure}
\caption{Nucleon plus-momentum distribution function, $f_N(y)$, for
$^{40}$Ca. Here $y\equiv p^+/(M_A/A)$}
\label{fig:fny}
\end{figure}
\begin{figure}
\caption{Vector meson plus-momentum distribution, $y f_v(y)$, for $^{40}$Ca.} 
\label{fig:fvy}
\end{figure}

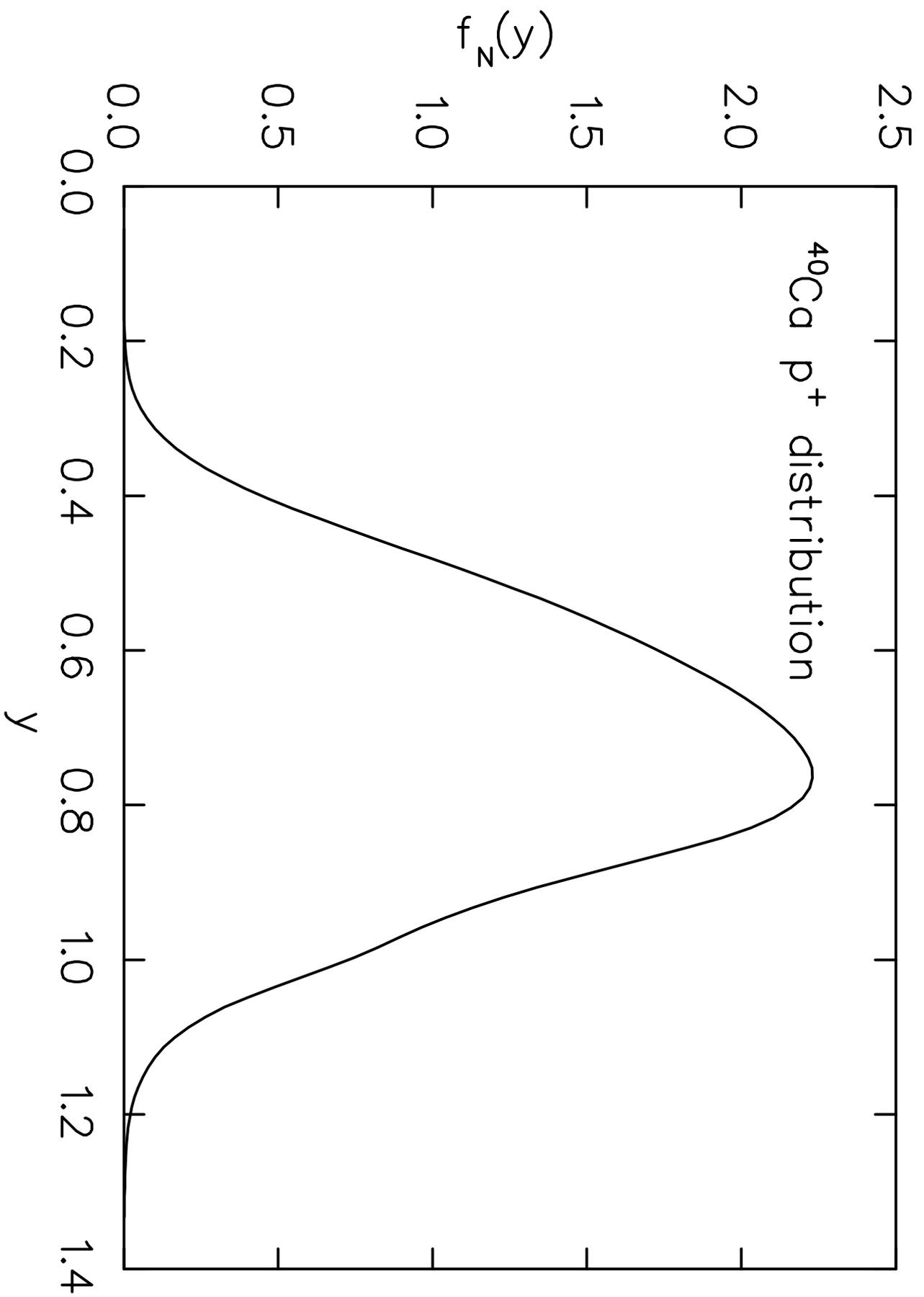
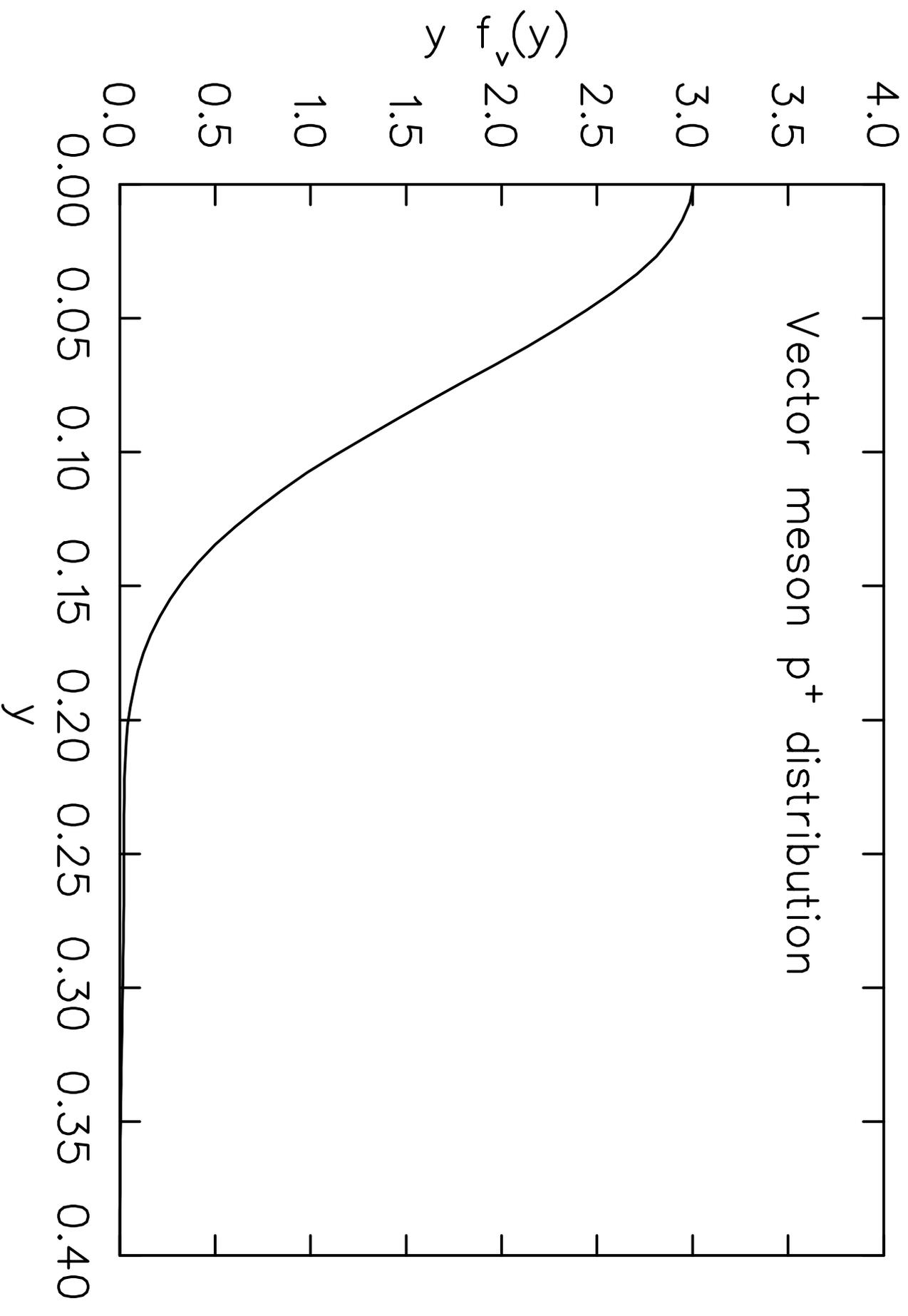
\end{document}